\documentclass[showpacs, pra,onecolumn,preprintnumbers ,amsmath, amssymb, superscriptaddress, aps]{revtex4-2}
\usepackage{color}
\usepackage{amsmath,amssymb}
\usepackage{pifont}
\usepackage{amssymb}  
\usepackage{bbold}
\usepackage{float}
\usepackage{subfloat}

\usepackage[caption=false]{subfig}
\usepackage{tikz}
\usepackage{makecell}
\usepackage{subfig}
\usepackage{pifont}   
\usepackage{graphicx} 
\graphicspath{{Figures/}}
\usepackage{dcolumn}  
\usepackage{bm}       
\usepackage{multirow} 
\usepackage{placeins}
\usepackage[colorlinks]{hyperref}
\usepackage{mathtools}
\usepackage{appendix}

\def \be{\begin{align}}
	\def \ee{\end{align}}
\def \bea{\begin{eqnarray}}
	\def \eea{\end{eqnarray}}

\begin{document}
	
	\title{Transmission in graphene through a double laser barrier}
	
	\author{Rachid El Aitouni}
	\affiliation{Laboratory of Theoretical Physics, Faculty of Sciences, Choua\"ib Doukkali University, PO Box 20, 24000 El Jadida, Morocco}
	
	\author{Miloud Mekkaoui}
	\affiliation{Laboratory of Theoretical Physics, Faculty of Sciences, Choua\"ib Doukkali University, PO Box 20, 24000 El Jadida, Morocco}
	\author{Ahmed Jellal}
	\email{ahmed.jellal@gmail.com}
	\affiliation{Laboratory of Theoretical Physics, Faculty of Sciences, Choua\"ib Doukkali University, PO Box 20, 24000 El Jadida, Morocco}
	\affiliation{Canadian Quantum  Research Center,
		204-3002 32 Ave Vernon,  BC V1T 2L7,  Canada}
	\begin{abstract}

		We study the tunneling behavior of Dirac fermions in graphene subjected to a double barrier potential profile created by spatially overlapping laser fields. By modulating the graphene sheet with an oscillating structure formed from two laser barriers, we aim to understand how the transmission of Dirac fermions is influenced by such a light-induced electric potential landscape.  Using the Floquet method, we determine the eigenspinors of the five regions defined by the barriers applied to the graphene sheet. Applying the continuity of the eigenspinors at barrier edges and using the transfer matrix method, we establish the transmission coefficients. These allow us to show that oscillating laser fields generate multiple transmission modes, including zero-photon transmission aligned with the central band $\varepsilon$ and photon-assisted transmission at sidebands  $\varepsilon+ l\varpi$, with $l=0,\pm1, \cdots$ and frequency $\varpi$. 
		For numerical purposes, our attention is specifically directed towards transmissions related to zero-photon processes ($l=0$), along with processes involving photon emission ($l=1$) and absorption ($l=-1$).
	We find that transmission occurs only when the incident energy is above the threshold energy $\varepsilon>k_y+2\varpi$, {with transverse wave vector $k_y$}. We find that the variation in distance {$d_1$ separating two barriers  of  widths $d_2-d_1$} suppresses one transmission mode. Additionally, we show that an increase in laser intensity modifies transmission sharpness and amplitude. 

	\end{abstract}

	\maketitle

\section{Introduction}

Graphene is a one-atom thick, two-dimensional carbon material in which carbon atoms are arranged in a hexagonal honeycomb lattice \cite{carbone,Novoselov}. It was first isolated in 2004 \cite{Novoselov2004} and has since demonstrated remarkable electronic and optical properties \cite{prop}. Graphene exhibits massless Dirac fermion behavior near the Fermi level \cite{masless} and the quantum Hall effect \cite{hall,Hall}, with charge carriers achieving high mobilities \cite{mobil,mobil2} approaching 1/300 the speed of light.
In addition, graphene possesses attractive photonic attributes such as broadband absorption of approximately $2.3\% $ of incident ultraviolet and infrared radiation \cite{absor}. Its properties are described by tight-binding model \cite{Tight}, with the dispersion relation exhibiting linear bands near the Dirac points \cite{desp2}. This implies nearly frictionless transport, as fermions readily traverse the conduction and valence bands.
However, robust tunneling even at sub-barrier energies, known as the Klein tunneling effect, hindered graphene's integration into devices \cite{klien1,klien2}. Experiments later verified Klein tunneling, where normally incident fermions transmit through barriers regardless of width or potential \cite{klienexp}.

	 Strategies to modulate Dirac fermion flow, such as mitigating Klein tunneling through barrier designs, could facilitate graphene's incorporation into electronics. Several methods have been developed, among them we cite depositing the graphene sheet onto a substrate \cite{substrat}, doping it with another type of atom \cite{dopage}, deforming the sheet \cite{def1, def2}, and applying an external electric, magnetic, or laser field.
 Klein tunneling is consistently observed in simple barrier. However, by adding a second barrier, the number of transmission oscillations  increases due to the additional resonance caused by the quasi-confined states between the two barriers \cite{double,Nbarreirs,doublebarr}.
 Confining fermions with a magnetic barrier results in the quantization of the energy spectrum \cite{mag4}. Despite this confinement, these fermions still exhibit Klein tunneling, highlighting a persistent and notable phenomenon. Remarkably, for a sequence of $N$ magnetic barriers, the transparency of transmission increases progressively with the growing value of $N$
	\cite{Nbarr,Nmag}. The quantification of the energy spectrum can also occur when dealing with a time-oscillating barrier \cite{condition1, Ojeda-Collado2013, timepot, timebarrier, butker}, giving rise to the emergence of multiple transmission modes \cite{bis1, bis2}.
The process of laser irradiation on graphene induces photon exchange between the graphene fermions and the barrier (laser field). Notably, increasing the intensity of the laser field has been demonstrated to effectively diminish overall transmission \cite{bis1, bis2}. In a related context, it has been observed that irradiating graphene subjected to an inclined potential barrier results in null transmission \cite{Elaitouni2023A}, leading to the complete confinement of fermions. This phenomenon is identified as anti-Klein tunneling.	 
Furthermore, the introduction of a magnetic field, along with a laser field, allows for the suppression of processes with zero photon exchange and the activation of processes involving photon exchange \cite{Elaitouni2022}.

We investigate the transmission of Dirac fermions through double laser barrier structures incorporated into a graphene sheet. The modulation of graphene with nanoscale laser potentials serves as a focal point for understanding how Dirac fermion transport is influenced by the induced photon-assisted tunneling dynamics within the barrier configuration. This exploration of light-matter interactions at the nano-scale interface between photonics and electronics has the potential to facilitate future control of emergent phenomena in graphene and other van der Waals materials.
To achieve our research objectives, we initiate the analysis by determining the eigenspinors within each region. This is accomplished by solving the eigenvalue equation using the Floquet method \cite{Floquet}. Subsequently, we apply boundary conditions at the interfaces of the double barriers and use the transfer matrix approach to derive all transmission modes.
In our numerical simulations, we specifically focus on transmissions involving zero photon exchange ($l=0$), as well as photon-assisted processes encompassing emission ($l=1$) or absorption ($l=-1$) of photons. This selective focus on first-order sideband transition modes streamlines the computational treatment while still providing valuable insights into the intricate interplay between photon-assisted tunneling and Dirac fermion transport, modulated by the laser barrier structure. 
We demonstrate that successful transmission occurs when the energy satisfies the condition $\varepsilon>k_y+2 \varpi$. This implies that fermions need to surpass a certain energy threshold to traverse the barrier. The manipulation of the distance between the two barriers and the adjustment of the barrier width allow for the modification of the amplitudes of two transmission modes or the suppression of one.
Moreover, the number of transmission oscillations increases across all regions, attributed to additional resonances arising from quasi-confined states between the two barriers. We illustrate that the intensity of the laser field directly influences both the amplitude of transmission and the overall transmission process. Importantly, our findings reveal that, irrespective of amplitude, the transmission modes exhibit sinusoidal variations.

The paper is organized  as follows: In Sec. \ref{TTMM}, we establish a theoretical model describing the system, which consists of five regions. We solve the eigenvalue equation to determine the eigenspinors associated with each region. In Appendix \ref{AA}, we apply the boundary conditions and utilize the current density to derive all transmission modes using the transfer matrix approach. By focusing on the first three modes, we present our numerical results and discuss the fundamental features of the system in Sec. \ref{TrMo}. We then conclude our findings in Sec. \ref{CCC}.

\section{THEORETICAL MODEl}\label{TTMM}

\begin{figure}[H]
	\centering
	\includegraphics[scale=1]{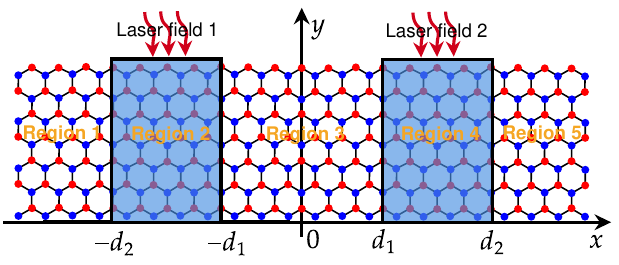}
	\caption{(Color online) {A diagram illustrates a graphene sheet with two laser barriers, each of width $d_2-d_1$, creating a system divided into five distinct regions.}}\label{fig1}
\end{figure}

We consider a graphene sheet partitioned into five regions, denoted by $j=1, 2, \cdots, 5$. In regions $1$, $3$, and $5$, pristine graphene is maintained. However, in regions $2$ and $4$, we introduce two distinct laser fields characterized by amplitudes $A_2$ and $A_4$, respectively, and phase-shifted by an angle $\beta$, as depicted in Fig. \ref{fig1}.
We want to highlight that our portrayal of the laser field as a step-like potential with nanometer resolution is a deliberately simplified modeling approach chosen for analytical tractability. This intentional simplification aims to capture essential aspects of the interaction while recognizing its inherently idealized nature. Our objective is not to precisely replicate real-world laser fields at the atomic scale but rather to establish a theoretical framework for examining the impact of double laser barriers on transmission in graphene. While our model provides valuable conceptual insights, we acknowledge the necessity for a more careful consideration of practical limitations.	
It is important to recognize that the idealization of perfectly rectangular laser barriers with nanometer resolution surpasses current experimental capabilities. Nevertheless, the scientific community is experiencing the development of innovative nanophotonics techniques that have the potential to shape optical fields at relevant length scales. In particular, progress in plasmonic nanostructures and metasurfaces presents promising avenues for localizing light in subwavelength regions through surface plasmons.
	Describing a laser field applied to graphene in the context of a step-like potential with nanometer resolution and modulating it at the same scale requires a multifaceted approach. Theoretical exploration entails utilizing quantum mechanical models, such as tight-binding or density functional theory, to gain a comprehensive understanding of the electronic behavior of graphene under the influence of a laser field \cite{carbone}. The notion of a step-like potential in laser-graphene interactions is based on the potential landscape experienced by electrons in the graphene lattice during laser irradiation, driven by the transfer of energy from the laser field \cite{2000}.
	To achieve nanometer resolution in describing a laser field on graphene, sophisticated experimental techniques become paramount. Instruments such as scanning tunneling microscopy (STM) and atomic force microscopy (AFM) enable researchers to investigate electronic and structural changes at the nanoscale with unprecedented precision \cite{3000}. Furthermore, modulating a laser field at the nanoscale on graphene entails external control mechanisms and dynamic tuning. Techniques like plasmonic modulation and the application of tailored external fields offer effective means to influence the laser-graphene interaction at the nanometer scale \cite{4000}.
	In summary, the comprehensive description of a laser field applied to graphene, considering both step-like potential changes and nanometer-scale modulation, requires an integrated approach that combines theoretical insights, advanced experimental techniques, and external control mechanisms. This holistic strategy is essential for advancing our understanding of laser-graphene interactions and their potential applications.

We emphasize that graphene is a highly relevant material due to its exceptional electronic, mechanical, and thermal properties \cite{carbone}. However, it is not the only relevant system in the field of materials science and condensed matter physics. Researchers have explored a wide range of two-dimensional materials and nanostructures, each possessing unique properties and potential applications.
	Other relevant systems include transition metal dichalcogenides (TMDs) like molybdenum disulfide (MoS$_2$) and tungsten diselenide (WSe$_2$)
	\cite{MakPRL2010, WangRMP2012}, as well as other carbon allotropes like carbon nanotubes and fullerenes \cite{IijimaN1991}. These materials exhibit distinct characteristics that make them suitable for various technological advancements.

To conduct a theoretical analysis of the present system, we construct a Hamiltonian that incorporates all the mentioned constraints. It can be expressed as follows:
\begin{equation}\label{Ham1}
	H_j=v_F\vec{\sigma}\cdot \left(\vec{p}+\frac{e}{c}\vec{A}_j(t)\right), \quad j=1,2,\cdots, 5
\end{equation}
 where  $v_F$ is the Fermi velocity, $\vec\sigma$ are the Pauli matrices $(\sigma_x,\sigma_y)$, and $\vec p=(p_x,p_y)$ is the momentum.  In the electric dipole approximation \cite{dipole}, $\vec{A}_j(t)$ represents the vector potential corresponding to the laser field
\begin{equation}
	\vec{A}_j(t)=(0,A_j \cos\Phi_j,0)
\end{equation}	
with $\Phi_j$ and $A_j$ being the phase and amplitude of the laser field in each region
\begin{align}
\Phi_j=\left\{\begin{array}{l}
	\Phi_1=0 \\
	\Phi_2= \omega t \\
	\Phi_3=0 \\
	\Phi_4= \omega t+\beta \\
	\Phi_5=0
\end{array},\quad A_j=\left\{\begin{array}{l}
	A_1=0, \ \ \ x<-d_2\\
	A_2, \ \ \ \ \ \ \ \ -d_2<x<-d_1 \\
	A_3=0, \ \ \ -d_1<x<d_1\\
	A_4, \ \ \ \ \ \ \ \ d_1<x<d_2 \\
	A_5=0, \ \ \ x>d_2
\end{array}\right.\right.
\end{align}
while the parameter  $\beta$ denotes a phase shift. This potential generates the oscillating electric field  $\vec{F}_j(t)=F_j\sin\Phi_j\ \vec{u}$ of  amplitude $F_j=\omega A_j$ and $\vec{u}$ is the unit vector along the polarization direction of the field (the $y$-axis). 

To determine the wave function corresponding to each region, we solve the wave equation by spiting the Hamiltonian \eqref{Ham1} into  spatial $H_0$ and  temporal $\widetilde{H}_j$ parts, i.e., $H_j=H_0+\widetilde{H}_j$, with
\begin{align}
	H_0=v_F  \left(\sigma_x p_x+\sigma_y p_y\right), \quad \widetilde{H}_j=v_F \sigma_y A_j(t).
\end{align}
Given that $H_0$ is time-independent and $\widetilde{H}_j$ is coordinate-independent, the total eigenspinors $	\Psi_j(x,y,t)=\psi_j(x,y)\phi_j(t) $ can be represented as a tensor product of the two eigenvectors   $\psi_j(x,y)=\dbinom{\varphi_{j}^+(x,y)}{\varphi_{j}^-(x,y)}$ and $\phi_j(t)$ associated with  $H_0$ and $\widetilde{H}_j$, respectively.
In the framework of the Floquet approximation \cite{bessel,floq}, the temporal part is expressed as $    \phi_j(t)=\chi_j(t)e^{-i\varepsilon t}$ (in system unit $\hbar=e=c=1$) associated with the Floquet energy $\varepsilon=\frac{E}{v_F}$, with $\chi_j(t)$ being a periodic function over time. The eigenvalue equation $H_j\Psi_j(x,y,t)=E\Psi_j(x,y,t)$  yields
\begin{align}
&\label{eq8}	\left[\partial_x +k_y-A_j \cos\Phi_j\right] \varphi_{j}^-(x,y)\chi_j(t)=\varphi_{j}^+(x,y)\frac{\partial}{\partial t}\chi_j(t)\\
&\label{eq9} 	\left[\partial_x -k_y+	A_j \cos\Phi_j\right]  \varphi_{j}^+(x,y)\chi_j(t)=\varphi_{j}^-(x,y)\frac{\partial}{\partial t}\chi_j(t).
\end{align}
{Unfortunately, It is not possible to determine the three unknown functions, ($\varphi_j^+, \varphi_j^-, \chi_j$), from these two equations. Thus, it is necessary to resort to some approximation. In this context, we use an iterative method to solve for the three variables \cite{bis2}. As a first approximation, we assume that $u_A(x)$ and $u_B(x)$ satisfy the coupled differential equations inside the barrier region without laser influence. Under this assumption, \eqref{eq8} and \eqref{eq9} to the following}
%
\begin{align}
	&\label{eq88}	-A_j \cos\Phi_j\  \varphi_{j}^-(x,y)\chi_j(t)=\varphi_{j}^+(x,y)\frac{\partial}{\partial t}\chi_j(t)\\
	&\label{eq99}	A_j \cos\Phi_j\  \varphi_{j}^+(x,y)\chi_j(t)=\varphi_{j}^-(x,y)\frac{\partial}{\partial t}\chi_j(t)
\end{align}
which gives rise to the following second order differential equation 
\begin{equation}
\left({\partial_t^2}+\omega \tan\Phi_j\ \partial_t+A_j^2\cos^2\Phi_j\right) \chi_j(t)=0
\end{equation}
having the solution
\begin{equation}
	\chi_j(t)=e^{-i\alpha\sin\Phi_j}=\sum_{m=-\infty}^{+\infty}J_m(\alpha_j)e^{-m\Phi_j}
\end{equation}
where $J_m$ is the Bessel functions, 
	and	$\alpha_j=\frac{F_j}{\omega^2}$.
Combing all to write the eigenspinors of $H_j$ as
\begin{equation}\label{sp2}
	\Psi_j(x,y,t)=\psi_j(x,y){ \sum_{m=-\infty}^{+\infty}}J_m(\alpha_j)e^{-i(\varepsilon t+m\Phi_j)}.
\end{equation}

To obtain a comprehensive derivation of \eqref{sp2}, it is necessary to determine $\psi_j(x,y)$. Indeed, 
in  regions 1, 3 and 5 there is only pristine graphene, then the corresponding  eigenspinors can be written as \cite{bessel,timebarrier,Ojeda-Collado2013}
\begin{eqnarray}
	\Psi_1(x,y,t)&=&{ \sum_{l,m=-\infty}^{+\infty}}\left[\begin{pmatrix}
		1\\
		\gamma_l
	\end{pmatrix}\delta_{m,0}e^{ik^0_xx}+r_l\begin{pmatrix}
	1\\
	-\gamma^*_l
\end{pmatrix}e^{-ik^l_xx}\right]e^{ik_yy}\delta_{m,l}e^{-iv_F(\varepsilon+m\varpi)t}\\
\Psi_3(x,y,t)&=&{ \sum_{l,m=-\infty}^{+\infty}}\left[c_{1l}
\begin{pmatrix}
	1\\
	\gamma_l
\end{pmatrix}e^{ik^l_xx}+c_{2l}\begin{pmatrix}
	1\\
	-\gamma^*_l
\end{pmatrix}
e^{-ik^l_xx}\right]e^{ik_yy}\delta_{m,l}e^{-iv_F(\varepsilon+m\varpi)t}\\
\Psi_5(x,y,t)&=&{\sum_{l,m=-\infty}^{+\infty}}\left[t_l
\begin{pmatrix}
	1\\
	\gamma_l
\end{pmatrix}e^{ik^l_xx}+\mathbb{0}_l\begin{pmatrix}
	1\\
	-\gamma^*_l
\end{pmatrix}
e^{-ik^l_xx}\right]e^{ik_yy}\delta_{m,l}e^{-iv_F(\varepsilon+m\varpi)t}
\end{eqnarray}
and the associated energy is
\begin{align}
	\varepsilon+l\varpi=s_l\sqrt{(k^l_x)^2+k_y^2}
\end{align} where $\varpi=\frac{\omega}{v_F}$, $\gamma_l=s_l\frac{k^l_x+ik_y}{\sqrt{(k^l_x)^2+k_y^2}}=s_le^{i\theta_l}$, $\theta_l=\arctan \frac{k_y}{k_x^l}$, $\delta_{m,l}=J_{m-l}(0)$,  $s_l=\text{sgn}(\varepsilon+l\varpi)$, $\mathbb{0}_l$ is the null vector, and $c_{il}$ $(i=1,2)$ are two constants. The coefficients $r_l$ and $t_l$ represent the reflection and transmission amplitudes, and their derivation will be provided in Appendix \ref{AA}.

In both regions 2 and 4, where laser fields are applied, the eigenvalue equation yields
\begin{eqnarray}
\left(-i\partial_x-i(k_y-m\varpi)\right)\varphi_{2j}(x,y)&=&(\varepsilon+m\varpi)\varphi_{1j}(x,y)\\
\left(-i\partial_x+i(k_y-m\varpi)\right)\varphi_{1j}(x,y)&=&(\varepsilon+m\varpi)\varphi_{2j}(x,y)
\end{eqnarray}
 which can be further analyzed to arrive at the solutions
\begin{eqnarray}
\Psi_2(x,y,t)&=&{ \sum_{l,m=-\infty}^{+\infty}}\left[a_{1l}\begin{pmatrix}
	1\\
\Gamma_l
\end{pmatrix}e^{iq^l_xx}+a_{2l}\begin{pmatrix}
	1\\
	-\Gamma^*_l
\end{pmatrix}e^{-iq^l_xx}\right]e^{ik_yy}J_{m-l}(\alpha_2)e^{-iv_F(\varepsilon+m\varpi)t}\\
\Psi_4(x,y,t)&=&{ \sum_{l,m=-\infty}^{+\infty}}\left[b_{1l}\begin{pmatrix}
	1\\
	\Gamma_l
\end{pmatrix}e^{iq^l_xx}+b_{2l}\begin{pmatrix}
	1\\
	-\Gamma^*_l
\end{pmatrix}e^{-iq^l_xx}\right]e^{ik_yy}J_{m-l}(\alpha_4)e^{-iv_F(\varepsilon+m\varpi)t}e^{-i(m-l)\beta}
\end{eqnarray}
and the corresponding  energy is 
\begin{align}
\varepsilon+l\varpi=s_l\sqrt{(q^l_x)^2+(k_y-l\varpi)^2}
\end{align}
where we have defined  $\Gamma_l=s_l\frac{q^l_x+i(k_y-l\omega)}{\sqrt{(q^l_x)^2+(k_y-l\varpi)^2}}=s_le^{i\theta_l'}$. The coefficients $a_{il}$ and $b_{il}$ $(i=1,2)$ are four constants. It is noteworthy that the $x$-component $q^l_x$ of the wave vector undergoes modification due to the laser field, in contrast to the scenario observed in the case of an oscillating barrier \cite{timepot,timebarrier,timebarrier1}.

\section{Transmission modes}\label{TrMo}

In Appendix {\color{red} A},
we have determined all transmission modes and the associated total transmission. Then, according to  \eqref{A30} and \eqref{A31}, we have
\begin{align}
&	T_m=\frac{\cos \theta_m}{\cos \theta_0}|t_m|^2 \\ &T=\sum_{m=-N}^{N}T_m
\end{align}
where $\cos \theta_m= \frac{k_x^m}{\sqrt{(k_x^m)^2+k_y^2}}$ and $k_x^m=\sqrt{(\varepsilon+m\varpi)^2-k_y^2}$.
	To gain a comprehensive understanding of how two time-oscillating laser barriers influence Dirac fermions in graphene, we numerically simulate our results. The oscillation of the barriers generates multiple energy bands, indicating infinite transmission modes. This includes zero-photon transmission corresponding to the central energy band $\varepsilon$, as well as photon-assisted transmission aligned with sidebands at $\varepsilon+l \varpi$. For simplifying the graphical representation, we focus on the first three bands: the central band and the first two sidebands. Numerical simulation allows probing transmission dynamics across these bands to elucidate the impact of tunneling and photon exchange processes on Dirac fermion transport induced by the double laser barrier structure.

\begin{figure}[!ht]
	\centering
	\subfloat[]{\centering\includegraphics[scale=0.68]{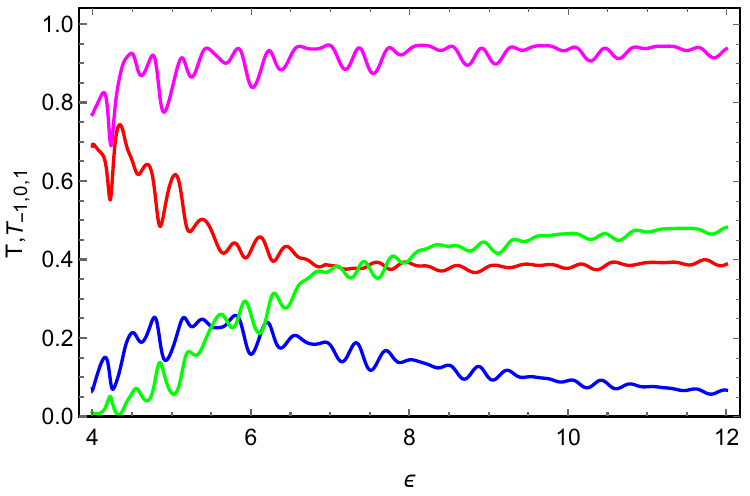}\label{fig2a}}\ \ \ \ \ \ \ \subfloat[]{\centering\includegraphics[scale=0.58]{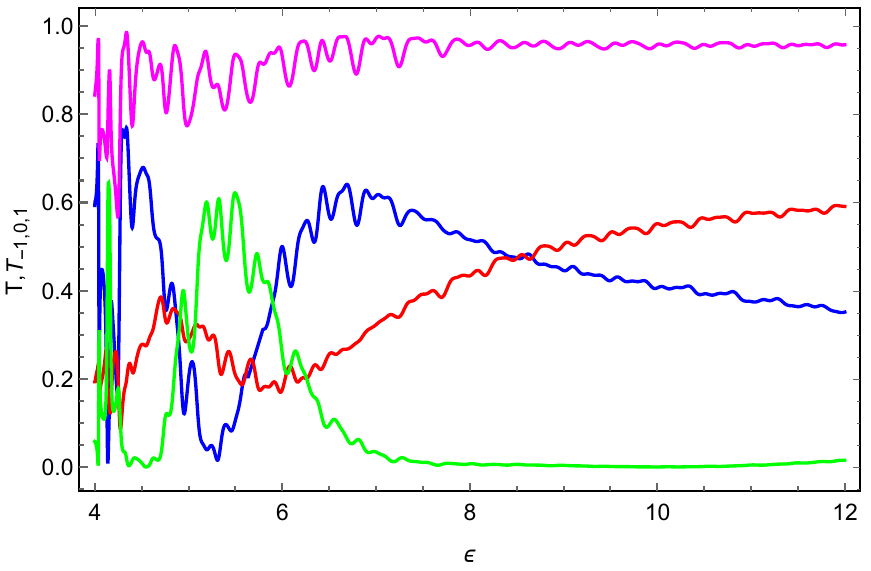}\label{fig2b}}\\
	\subfloat[]{\centering\includegraphics[scale=0.58]{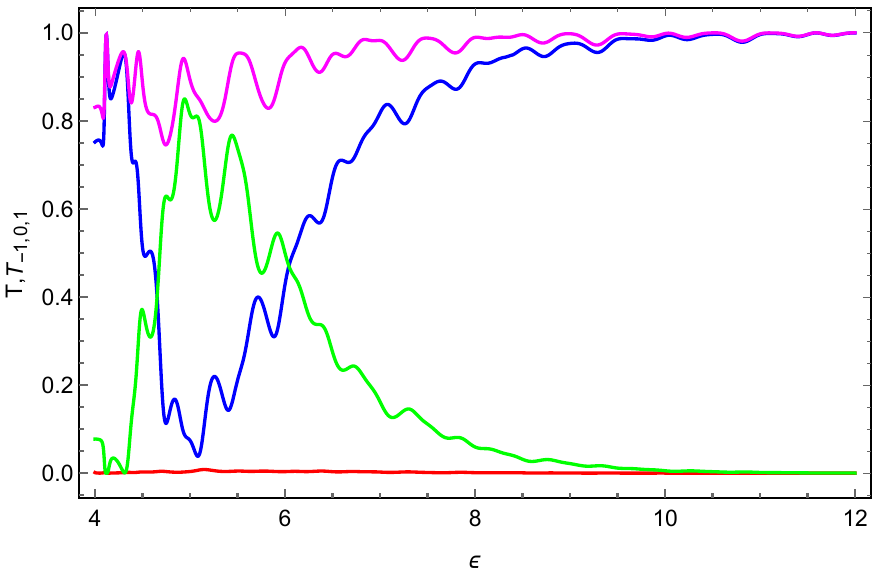}\label{fig2c}}\ \ \ \ \ \ \ \subfloat[]{\centering\includegraphics[scale=0.58]{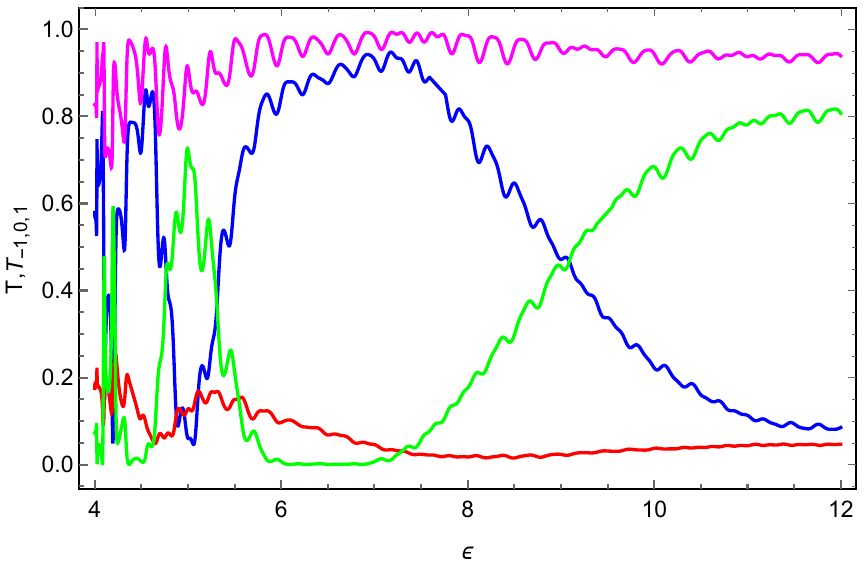}\label{fig2d}}
	\caption{{(Color online) Transmissions as a function of the  energy $\varepsilon$ for $\omega=1.5$, $k_y=1$, $\beta=\frac{\pi}{8}$, $F_2=F_4=1.5$, and different distances ($d_1,d_2$) with (a): $d_1=3$, $d_2=5$, (b): $d_1=3$, $d_2=10$, (c): $d_1=1$, $d_2=5$, (d): $d_1=1$, $d_2=10$. $T_1$ (red line),  $T_{-1}$ (green line), $T_{0}$  (blue line), and $T$ (magenta line).}}\label{fig2}
\end{figure}

Fig. \ref{fig2} depicts the transmissions as a function of the energy $\varepsilon$ for various values of {$d_1$ and $d_2$, where $d_1$ is the distance between the two barriers of  width $d_2-d_1$}.
We take into account the condition $\varepsilon>k_y+2\varpi$ in order to have transmission, and we mention that the quantity $k_y+2\varpi$ serves as an effective mass \cite{masse}. 
We observe that the transmissions vary in an oscillatory way, and the total transmission oscillates in the vicinity of unity. Fig. \ref{fig2a} plotted for $d_1=3$ and $d_2=5$, we see that the transmission with photon emission $T_1$ (red line) decreases exponentially, the transmission with absorption $T_{-1}$ (green line) increases along the $\varepsilon$-axis, and the transmission $T_{0}$ with zero photon exchange (blue line) decreases rapidly towards zero. When $d_2$ is increased,  $T_{-1}$ becomes null from $\varepsilon=7$ but $T_1$ increases, and $T_{0}$ shows different behaviors varying between decreasing then increasing along the $\varepsilon-$axis as depicted in Fig. \ref{fig2b}. 
For the values $d_1=1$ and $d_2=5$, $T_1$ is almost zero for all  energies,  $T_{-1}$ increases then decreases exponentially, and  $T_{0}$ shows the opposite behavior (decreases then increases exponentially). From the value $\varepsilon=10$, we see that  the transmission is carried out only with zero photon exchange ($T_{0}$) and then all  fermions cross the barrier, showing {perfect transmission.}
For $d_2=10$ in Fig. \ref{fig2d}, we observe that $T_{0}$ is dominant between $\varepsilon=6$ and $\varepsilon=9$, and after that,  $T_{1}$ becomes more dominant but $T_{-1}$ is almost null.
In conclusion, the modulation of the two distances $d_1$ and $d_2$ serves as a means to control transmission modes, consistent with observations for time-oscillating barriers \cite{Ojeda-Collado2013}. Notably, adjusting one of the two distances can influence transmission properties. However, it is important to highlight that complete suppression of Klein tunneling is not achievable in graphene. Overall, the separation between laser-induced barriers emerges as a design parameter offering tunable transmission through the graphene sheet, with Klein tunneling persisting as an underlying effect shaped by the massless nature of Dirac fermions.

\begin{figure}[!ht]
	\centering
	\subfloat[]{\centering\includegraphics[scale=0.65]{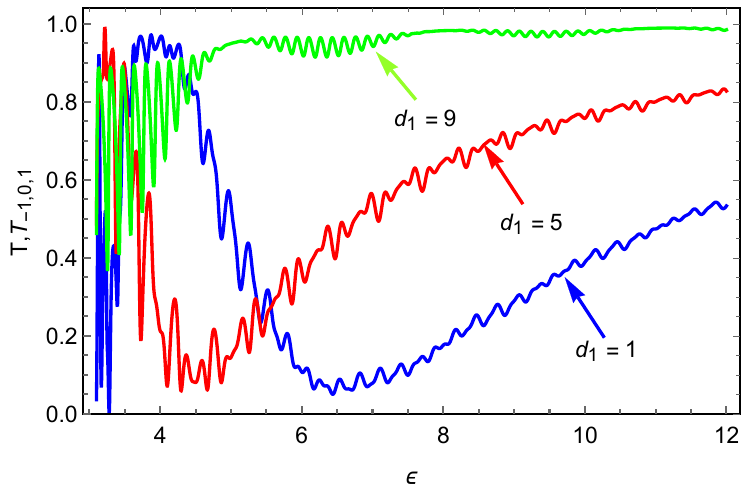}\label{fig3a}}\ \ \ \ \ \ \ \
	\subfloat[]{\centering\includegraphics[scale=0.65]{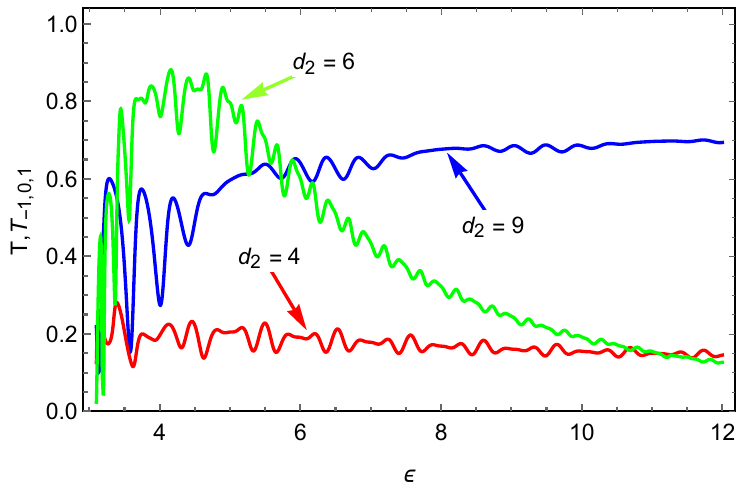}\label{fig3b}}\\
	\caption{{(Color online) Transmission $T_0$ with zero photon exchange  as a function of the energy $\varepsilon$  for $\beta=0$, $k_y=0.1$, $\varpi=1.5$, $F_2=F_4=1.5$, and different  distances ($d_1,d_2$) with (a): $d_2=10$, $d_1=1$ (blue line), $d_1=5$ (red line), $d_1=9$ (green line), and (b): $d_1=3$, $d_2=4$ (blue line), ${d_2=6}$ (red line), ${d_2=9}$ (green line).}}\label{fig3}
\end{figure}
In Fig. \ref{fig3}, we  plot the transmission $T_0$ of the central band ($l=0$) as a function of the  energy $\varepsilon$ for different values of $d_1$ and $d_2$. Fig. \ref{fig3a} illustrates  the  behavior of $T_0$ for three values $d_1=1,5,9$. It is evident that as $d_1$ increases, the number of oscillations also increases, and $T_{0}$ becomes more dominant in the contribution to the total transmission because the two barriers resemble two peaks of width $d=d_2-d_1$. In this scenario, a majority of incident fermions traverse the barrier with zero photon exchange, as evident in the green curve corresponding to $d_1=9$. Now for $d_2=2,6,9$, we present $T_0$ in Fig. \ref{fig3b}. We observe that for $d_2=4$ (red line), $T_{0}$ is very weak and decreases in an oscillatory way. For $d_2=6$ (green line), $T_0$ increases for low energies, then decreases exponentially towards zero in the vicinity of $\varepsilon=11$. For $d_2=9$ (blue line), $T_{0}$ becomes more important than the other transmission process because the sum of the three transmission modes is close to unity. We can conclude that increasing the width of barriers {\color{red}d} suppresses the transmission of the side bands and increases the transmission with zero photon exchange.
 
 \begin{figure}[!ht]
 	\subfloat[]{\centering\includegraphics[scale=0.65]{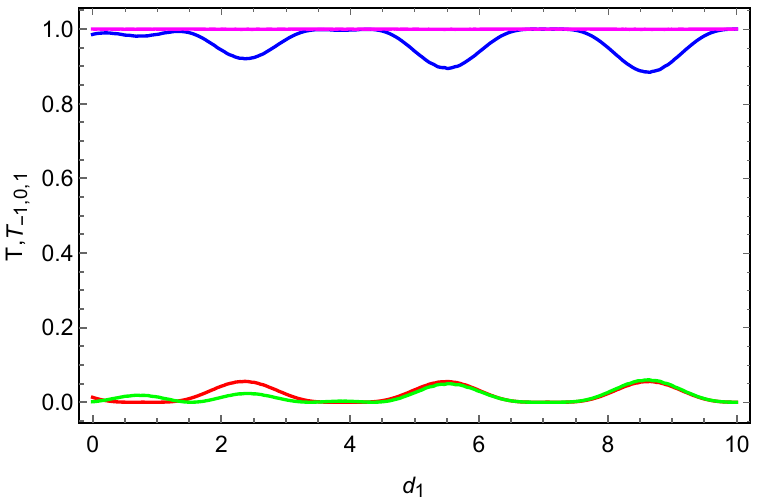}\label{fig4a}}\ \ \ \ \subfloat[]{\centering\includegraphics[scale=0.65]{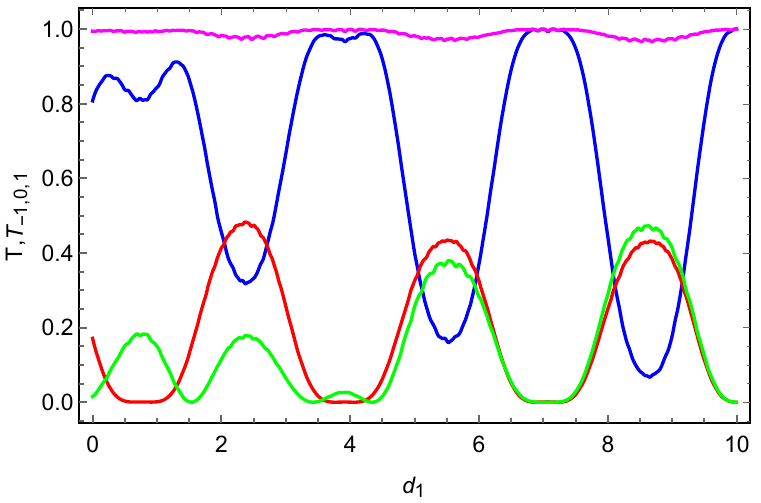}\label{fig4b}}\\
 	\subfloat[]{\centering\includegraphics[scale=0.65]{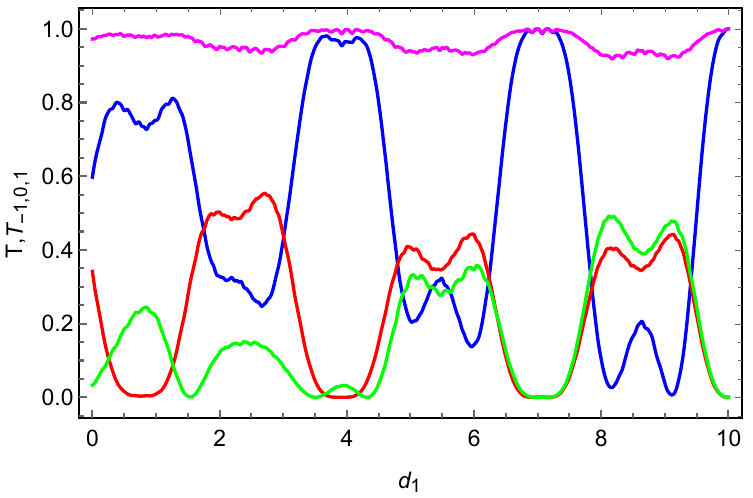}\label{fig4c}}\ \ \ \
 	\subfloat[]{\centering\includegraphics[scale=0.65]{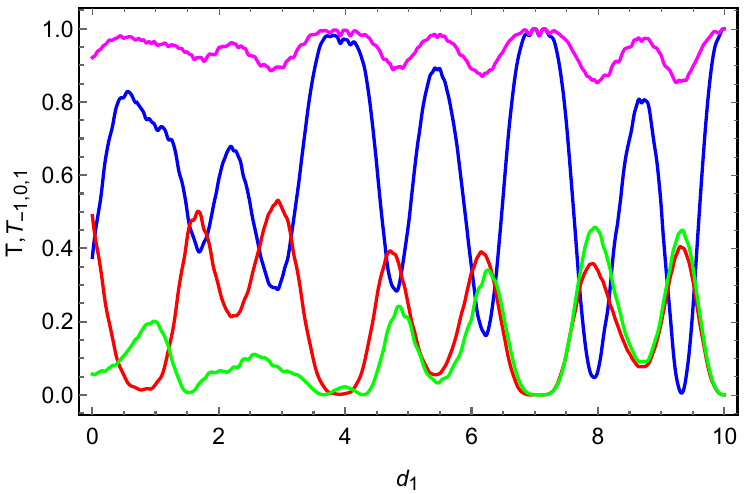}\label{fig4d}}
 	\caption{{(Color online) Transmissions as a function of the distance $d_1$, separating two barriers,  for $d_2=10$, $k_y=1$, $\varpi=2$, $\varepsilon=20$, $\beta=0$,
 			and  different amplitudes $F$ ($F_1=F_4=F$) with
 			(a): $F=0.5$, (b): $F=1.9$, (c): $F=2.9$, (d): $F=3.9$. With
 			$T$ (magenta line), $T_0$ (blue line), $T_1$ (red line), $T_{-1}$ (green line).}}\label{fig4}
 \end{figure}

In Fig. \ref{fig4}, we present the transmissions as a function of the distance $d_1$ between the two barriers for different values of the amplitude $F$ of the laser field. For $F=0.5$ in Fig. \ref{fig4a}, we observe that the total transmission $T$ (magenta line) almost equals unit whatever the distance $d_1$ because the laser fields are very weak and they have almost a negligible effect. The transmission $T_0$ with zero photon exchange oscillates in the vicinity of the unit, and the transmissions ($T_1$, $T_{-1}$) with photon exchange oscillate in the vicinity of zero. The laser fields are very weak but allow for quantifying the energy, even though the majority of the fermions cross the barrier with zero photon exchange. 
For $F=1.9$ in Fig. \ref{fig4b}, we see that the laser effect is very clear because $T_0$ varies periodically, $T_1$ (red line) decreases, and $T_ {-1}$ increases along the $d_1$-axis. However, $T_0$ varies in phase opposition with the two other transmission modes, with an increase in the amplitude of the oscillations along the $d_1$-axis. {The total transmission remains consistently close to unity, demonstrating the transparency of two barriers}.
{indicating barrier transparency.}
For $F=2.9$ in Figure \ref{fig4c}, $T_0$ exhibits oscillations with an amplification in amplitude, and it diminishes for  specific values. 
For $F=3.9$ in Fig. \ref{fig4d}, a reduction in the interval where $T_{1}$ and $T_{-1}$ are canceled is observed, accompanied by an increase in the number of peaks. It is noticeable that $T_1$ oscillates with a decrease in amplitude while $T_{-1}$ increases. Consequently, increasing the amplitude $F$ of the laser fields seems to reduce the interval of transmission cancellation, but it also leads to an increase in the number of oscillations.

\begin{figure}[!ht]
	\subfloat[]{\centering\includegraphics[scale=0.6]{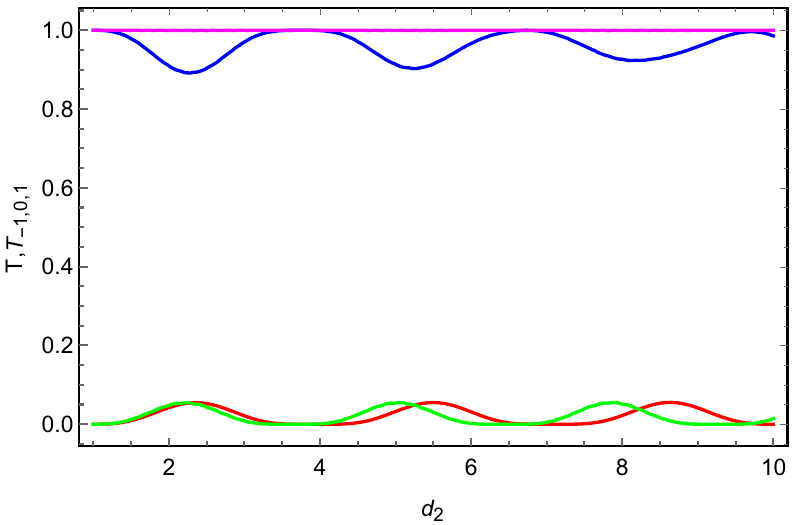}\label{fig5a}}\ \ \ \ \ \ \ \ \subfloat[]{\centering\includegraphics[scale=0.6]{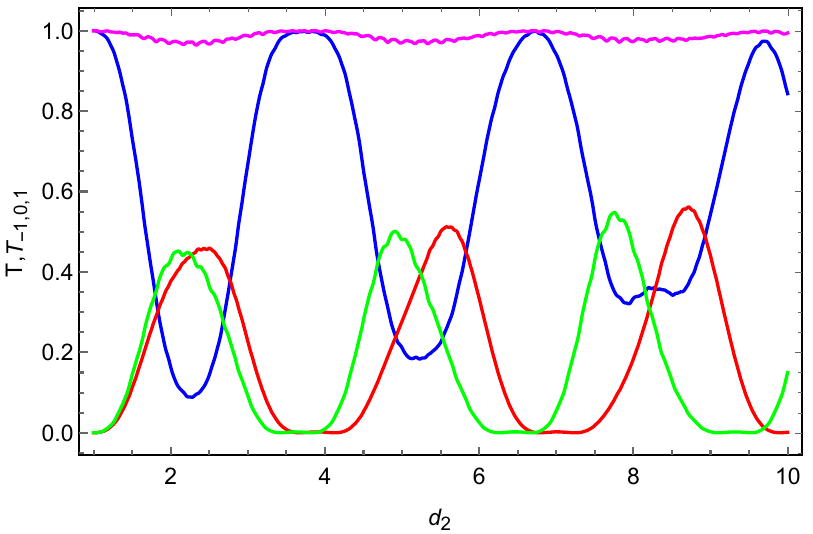}\label{fig5b}}\\
	\subfloat[]{\centering\includegraphics[scale=0.6]{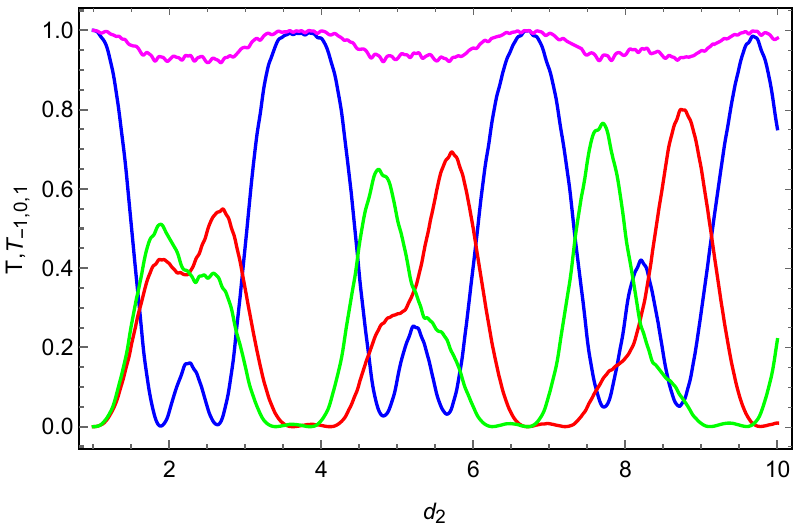}\label{fig5c}}\ \ \ \ \ \ \ \ 
	\subfloat[]{\centering\includegraphics[scale=0.6]{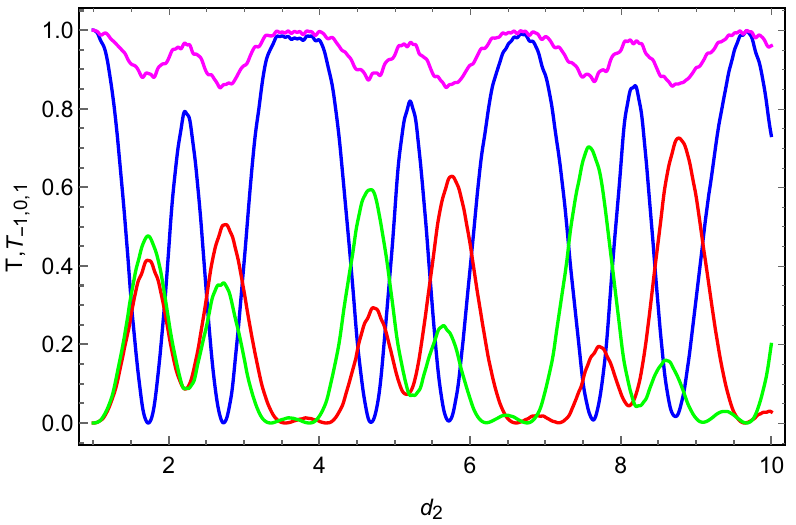}\label{fig5d}}
	\caption{{(Color online) Transmissions as a function of the distance $d_2$, 
			 for ${d_1=1}$, $k_y=1$, $\varpi=2$, $\varepsilon=20$, $\beta=0$,
			and  different amplitudes $F$ ($F_2=F_4=F$) with
			(a): $F=0.5$, (b): $F=1.9$, (c): $F=2.9$, (d): $F=3.9$.  With
			$T$ (magenta line), $T_0$ (blue line), $T_1$ (red line), $T_{-1}$ (green line).}}\label{fig5}
\end{figure}

Fig. \ref{fig5} presents transmissions as a function of {the distance $d_2$ for a fixed $d_1$, where $d_2-d_1$ is the width of two barriers}. 
for the same configuration of parameters taken in Fig. \ref{fig4}. For $F=0.5$ in Fig. \ref{fig5a}, we observe a generation of transmission modes, but the effect of the laser fields is very weak. The transmissions $T_{1}$ and $T_{-1}$ oscillate around zero, which can be neglected compared to the contribution of $T_{0}$, implying 
{perfect transmission.} For  $F=1.9$ in Fig. \ref{fig5b}, we notice that $T_{1}$ and $T_{-1}$ vary periodically with an increase in amplitude, but $T_0$ varies in an oscillatory way with a decrease in the amplitude. For $F=2.9$, Fig. \ref{fig5c} shows that $T_0$ varies regularly with an appearance of peaks in the minimum part. For $F=3.9$ in Fig. \ref{fig5d}, we observe that the total transmission $T$ oscillates around the unit, while $T_0$ is more dominant, oscillating between zero and one, canceling out at several points. $T_{1}$ and $T_{-1}$ also oscillate with an increase in amplitude. For example, in the vicinity of the value $d_2=4$, $T_{1}$ and $T_{-1}$ are almost null, but $T_0$ is equal to the unit, which implies that all the fermions cross the barrier without exchanging photons.

\begin{figure}[!ht]
	\subfloat[]{\centering\includegraphics[scale=0.6]{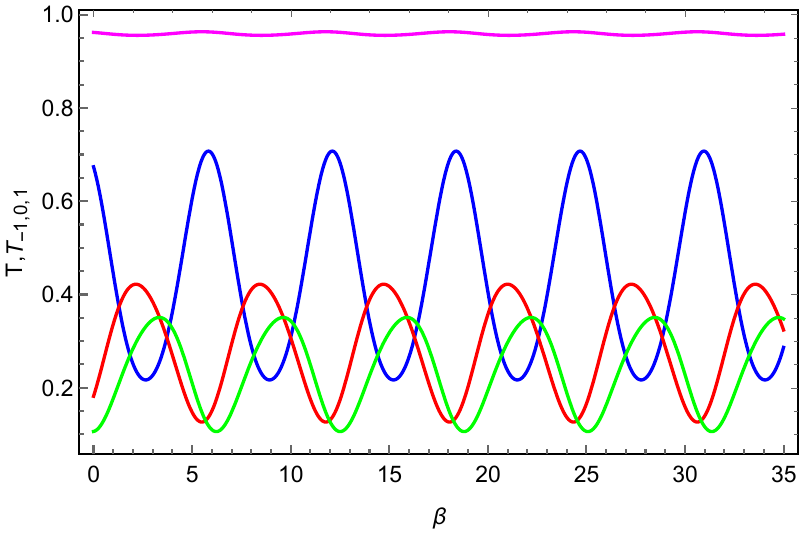}\label{fig6a}}\ \ \ \ \ \ \ \ \subfloat[]{\centering\includegraphics[scale=0.6]{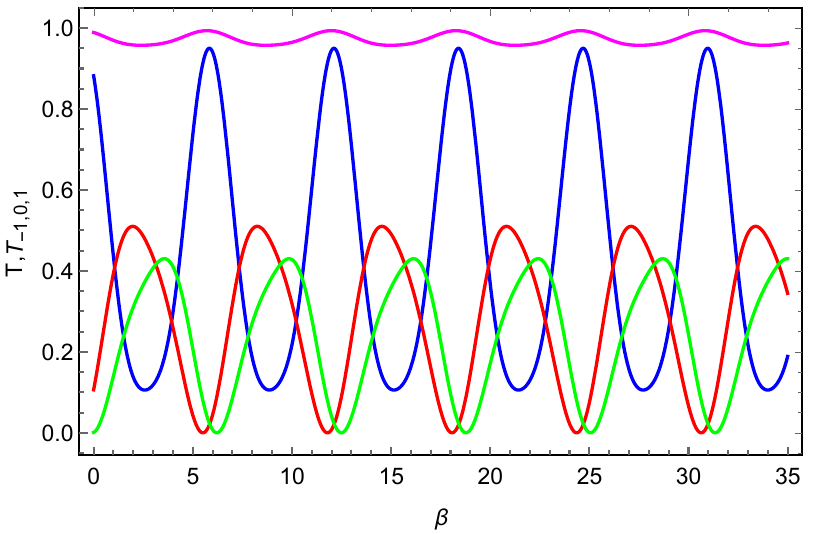}\label{fig6b}}\\
	\subfloat[]{\centering\includegraphics[scale=0.6]{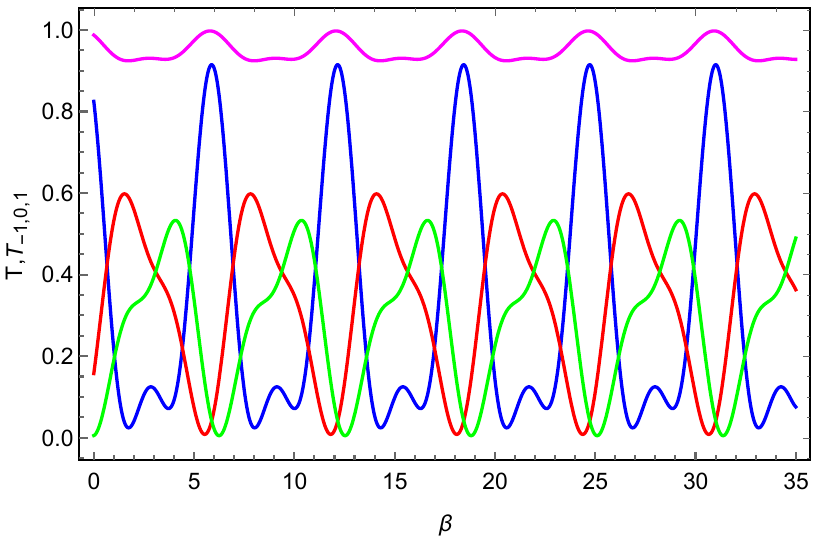}\label{fig6c}}\ \ \ \ \ \ \ \ \subfloat[]{\centering\includegraphics[scale=0.6]{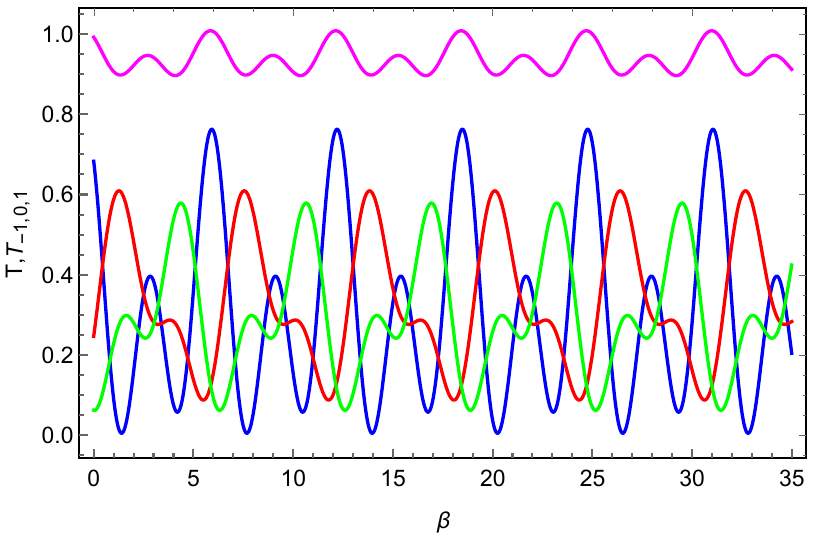}\label{fig6d}}
		
\caption{{(Color online) Transmissions as a function of the phase shift $\beta$  for $k_y=1$, $\varpi=2$, $\varepsilon=12$,  $F_2=2$, $d_1=1.3${\color{red},} $d_2=3$, and different value of amplitude  $F_4$ such that  (a): $F_4=0.9$, (b): $F_4=2$, (c): $F_4=2.9$, (d): $F_4=3.9$. With $T$ (magenta line), $T_0$ (blue line), $T_1$ (red line), $T_{-1}$ (green line).}}\label{fig6}
\end{figure}

In Fig. \ref{fig6}, we show transmissions as a function of the phase shift $\beta$ for $k_y=1$, $\varpi=2$, $\varepsilon=12$, $F_2=2$, $d_1=1.3$, $d_2=3$, and various amplitudes $F_4$ of the second laser barrier. In Fig. \ref{fig6a} with $F_4=0.9$, we notice that the three transmission modes $(T_0, T_1, T_{-1})$ exhibit consistent oscillations with the same frequency across different amplitudes. Notably, $T_0$ (zero photon) appears more predominant, while $T_1$ (photon emission) and $T_{-1}$ (photon absorption) exhibit similar variations. These two transmissions periodically nullify for specific values of $\beta$, resembling observations in the case of time-oscillating barriers \cite{Ojeda-Collado2013}. The total transmission $T$ oscillates around unity, {and then two  barriers are transparent}. In Fig. \ref{fig6b} with $F_4=2$, we observe that the amplitude of oscillations gets intensified, while $T$ exhibits fluctuations around unity with a dominance of $T_0$. In Fig. \ref{fig6c} with $F_4=2.9$, transmissions vary periodically by losing the sinusoidal behavior, and even the periodicity gets lost for some particular values of $\beta$. In this setup, $T_0$ maintains a dominant presence, exhibiting an amplitude close to 0.9, while $T_1$ and $T_{-1}$ undergo symmetrical variations. For $F_4=3.9$, Fig. \ref{fig6d} illustrates a decrease in the amplitude of $T_0$ with a simultaneous increase in the number of oscillations, leading to periodic cancellations. The three transmissions become nearly equiprobable, 
{and the periodic observation of perfect transmission occurs.}

\section{Conclusion}\label{CCC}

We have studied the transmission of Dirac fermions through double laser barriers generated by two electric fields of amplitude $(F_2, F_4)$ and frequency $\omega$ shifted by a phase shift $\beta$. The two barriers divide the graphene sheet into five regions, such that in regions 1, 3, and 5, there is only pristine graphene, and the other two regions are irradiated by laser fields. By using the Floquet approximation, we analytically determined the eigenspinors associated with each region. Additionally, we showed that the oscillation of the barriers over time generates several energy bands denoted by $\varepsilon+ l\varpi$ with $l=0,\pm1, \cdots$. Subsequently, we have determined the transmission modes by applying the boundary conditions at the interfaces of the two barriers and using the transfer matrix method together with current density.

For numerical illustrations, we restricted ourselves to the three first modes
corresponding to  the central band $\varepsilon$ and the first two side bands correspond to the $\varepsilon\pm\varpi$. We showed that  the transmission exists if the incident energy of the Dirac fermions satisfies the condition $\varepsilon>k_y+ 2\varpi$.
It was observed that the time-varying barrier induces two transmission processes: one without and one with photon exchanges.
It was demonstrated that varying the distance $d_1$, which separates the two barriers, offers a means to cancel one of the two transmission processes, modify the number of oscillations, and adjust the transmission processes. It was observed that an increase in $d_1$ results in a higher number of oscillations. This is attributed to additional resonance between the barriers and the quasi-confined states, activating the transmission process with zero photon exchange while suppressing the transmission processes involving photon exchange.
Reducing the {distance $d_2$}
leads o a decrease in the transmission $T_0$ with zero photon exchange, even as the incident energy $\varepsilon$ increases. An increase in the laser field amplitude raises both the number of oscillations and their associated amplitude. Periodic observation of 
{perfect transmission}
 is noted, occurring for very precise values of the phase shift $\beta$.

This study serves as a crucial step in laying the groundwork for the development of optoelectronic technologies utilizing photon-assisted tunneling in graphene. By advancing our understanding of light-matter interactions at the nanoscale, the current research establishes a foundation for the future application of graphene in devices driven by transient optical fields. The insights gained from this study are poised to inspire further research aimed at harnessing photon-induced tunneling effects for diverse applications, including ultrafast optoelectronics, renewable energy generation, and quantum information systems that leverage the distinctive characteristics of Dirac fermions. In essence, this investigation contributes to enabling the translation of emerging nanophotonics concepts into new generations of innovative optoelectronic architectures, taking advantage of graphene's unique light-sensitive properties.

{In the following, we discuss how to design a realistic experiment for the interaction of graphene electrons with a double barrier of laser beams.	
	In real experimental scenarios, the generation of time-oscillating barriers requires the manipulation of external fields or potentials to alter the potential landscape experienced by particles or waves. Researchers typically use two methods for this purpose. The first method involves the use of an optical grating created by the interference of laser beams, resulting in the formation of a periodic potential \cite{3535, 3636}. By carefully controlling the laser parameters, such as intensity and phase, the optical lattice can be tailored to create the desired time-varying barriers. The second method involves the application of time-dependent electric fields. By subjecting a device, such as a semiconductor heterostructure or microfabricated device, to a varying voltage, an electric field is induced. This time-dependent electric field can be modulated in a sinusoidal manner or shaped with other time-dependent waveforms \cite{3737, 3838}. These variations in the electric field result in the formation of time-varying barriers, allowing researchers to study the interactions of particles or waves with this dynamic potential landscape.}

 \section*{Acknowledgment}
We thank Prof. A. H. Alhaidari for valuable discussions.

\section*{Conflict of interest statement}

 	The authors declare that they have no known competing financial interests or personal relationships that could have appeared to influence the work reported in this paper. 

\section*{Authors contribution statement}
All authors have contributed equally to the paper.

\section*{Data Availability Statement}
The data that support the findings of this study
are available on request from the corresponding author.

\appendix\label{Appendix}

\section{Transmission modes}\label{AA}

To derive the transmission modes, we use the continuity of the eigenspinors at the barrier interfaces. Indeed, 
for the first barrier, we write  $\Psi_1(-d_2,y,t)=\Psi_2(-d_2,y,t)$ and $\Psi_2(-d_1,y,t)=\Psi_3(-d_1,y,t)$, which allow us to get the set of equations
\begin{align}
	&\delta_{m,0}e^{-ik^0_xd_2}+r_me^{+ik^m_xd_2}=\sum_{l=-\infty}^{+\infty}\left(a_{1l}e^{-iq^l_xd_2}+a_{2l}e^{+iq^l_xd_2}\right)J_{m-l}(\alpha_2)\\
	&\delta_{m,0}\gamma_me^{-ik^0_xd_2}-r_m\gamma^*_me^{+ik^m_xd_2}=\sum_{l=-\infty}^{+\infty}\left(a_{1l}\Gamma_le^{-iq^l_xd_2}-a_{2l}\Gamma^*_le^{+iq^l_xd_2}\right)J_{m-l}(\alpha_2)\\
	&c_{1m}e^{-ik^m_xd_1}+c_{2m}e^{+ik^m_xd_1}=\sum_{l=-\infty}^{+\infty}\left(a_{1l}e^{-iq^l_xd_1}+a_{2l}e^{+iq^l_xd_1}\right)J_{m-l}(\alpha_2)\\
	&c_{1m}\gamma_me^{-ik^m_xd_1}-c_{2m}\gamma^*_me^{+ik^m_xd_1}=\sum_{l=-\infty}^{+\infty}\left(a_{1l}\Gamma_le^{-iq^l_xd_1}-a_{2l}\Gamma^*_le^{+iq^l_xd_1}\right)J_{m-l}(\alpha_2).
\end{align}
As for  the second barrier,  we have $\Psi_3(d_1,y,t)=\Psi_4(d_1,y,t)$ and $\Psi_4(d_2,y,t)=\Psi_5(d_2,y,t)$, then we end up with the following set
\begin{align}
	&c_{1m}e^{ik^m_xd_1}+c_{2m}e^{-ik^m_xd_1}=\sum_{l=-\infty}^{+\infty}\left(b_{1l}e^{iq^l_xd_1}+b_{2l}e^{-iq^l_xd_1}\right)e^{-i(m-l)\beta}J_{m-l}(\alpha_4)\\
	&c_{1m}\gamma_me^{ik^m_xd_1}-c_{2m}\gamma^*_me^{-ik^m_xd_1}=\sum_{l=-\infty}^{+\infty}\left(b_{1l}\Gamma_le^{iq^l_xd_1}-b_{2l}\Gamma^*_le^{-iq^l_xd_1}\right)e^{-i(m-l)\beta}J_{m-l}(\alpha_4)\\
	&t_me^{ik^m_xd_2}+\mathbb{0}_me^{-ik^m_xd_2}=\sum_{l=-\infty}^{+\infty}\left(b_{1l}e^{iq^l_xd_2}+b_{2l}e^{-iq^l_xd_2}\right)e^{-i(m-l)\beta}J_{m-l}(\alpha_4)\\
	&t_m\gamma_me^{ik^m_xd_2}-\mathbb{0}_m\gamma^*_me^{-ik^m_xd_2}=\sum_{l=-\infty}^{+\infty}\left(b_{1l}\Gamma_le^{iq^l_xd_2}-b_{2l}\Gamma^*_le^{-iq^l_xd_2}\right)e^{-i(m-l)\beta}J_{m-l}(\alpha_4).
\end{align}
The above  sets can be mapped in the  matrix form 
\begin{equation}\label{matrice}
	\begin{pmatrix}
		\delta_{m,0}\\
		r_m
	\end{pmatrix}
	=\begin{pmatrix}
		\mathbb{S}_{1,1}&	\mathbb{S}_{1,2}\\
		\mathbb{S}_{2,1}&	\mathbb{S}_{2,2}
	\end{pmatrix}	\begin{pmatrix}
		t_m\\
		\mathbb{0}_m
	\end{pmatrix}=\mathbb{S}
	\begin{pmatrix}
		t_m\\
		\mathbb{0}_m
	\end{pmatrix}
\end{equation}
where $\mathbb{S}$ is the  transfer matrix. It is essential to note that $\mathbb{S}$ has infinite order, and for simplification, we truncate it to finite order by considering $m$ between -$N$ and $N$, where $N>\alpha_j$ \cite{condition1, timepot}. $\mathbb{S}$ is expressed as a product of the transfer matrix $\mathbb{S}(j,j+1)$ from region $j$ to region $j+1$
\begin{equation}\label{trans}
	\mathbb{S}=\mathbb{S}(1,2).\mathbb{S}(2,3).\mathbb{S}(3,4).\mathbb{S}(4,5)
\end{equation} 
with different matrices
\begin{eqnarray}
	\mathbb{S}(1,2)&=&\begin{pmatrix}
		\mathbb{G}^0_{- d_{2}}& \mathbb{G}_{d_{2}} \\
		\mathbb{R}_{+,-d_{2}}^{+1}& \mathbb{R}_{-,d_{2}}^{-1}\\
	\end{pmatrix}^{-1}
	\begin{pmatrix}
		\mathbb{H}_{+,-d_2}^{2}& \mathbb{H}_{+,d_2}^{2}\\
		\mathbb{L}_{+,-d_2}^{2,+1}&\mathbb{L}_{-,d_2}^{2,-1}\\
	\end{pmatrix}\\
	\mathbb{S}(2,3)&=&
	\begin{pmatrix}
		\mathbb{H}_{+,-d_1}^{2}& \mathbb{H}_{+,d_1}^{2}\\
		\mathbb{L}_{+,-d_1}^{2,+1}&\mathbb{L}_{-,d_1}^{2,-1}
	\end{pmatrix}^{-1}
	\begin{pmatrix}
		\mathbb{G}_{- d_{1}}& \mathbb{G}_{d_{1}} \\
		\mathbb{R}_{+,-d_{1}}^{l,+1}& \mathbb{R}_{-,d_{1}}^{l,-1}\\
	\end{pmatrix}\\
	\mathbb{S}(3,4)&=&\begin{pmatrix}
		\mathbb{G}_{d_{1}}& \mathbb{G}_{-d_{1}}\\
		\mathbb{R}_{+,d_{1}}^{+1}& \mathbb{R}_{-,-d_{1}}^{-1}\\
	\end{pmatrix}^{-1}
	\begin{pmatrix}
		\mathbb{K}^4_{+,-d_1}& \mathbb{K}^4_{+,d_1}\\
		\mathbb{F}^{4,+}_{+,d_1}&\mathbb{F}^{4,+}_{-,-d_1}\\
	\end{pmatrix}\\
	\mathbb{S}(4,5)&=&
	\begin{pmatrix}
		\mathbb{K}^4_{+,d_2}& \mathbb{K}^4_{+,-d_2}\\
		\mathbb{F}^{4,+}_{+,d_2}&\mathbb{F}^{4,+}_{-,-d_2}\\
	\end{pmatrix}^{-1}
	\begin{pmatrix}
		\mathbb{G}_{d_{2}}& \mathbb{G}_{-d_{2}} \\
		\mathbb{R}_{+,d_{2}}^{+1}& \mathbb{R}_{-,-d_{2}}^{-1}\\
	\end{pmatrix}
\end{eqnarray}
where $z=\pm d_{1,2}$ and  we have set 
\begin{align}
	&(\mathbb{G}^{0}_{z})=e^{ik^0_xz}\\
	&(\mathbb{G}_{z})_{m,l}=e^{ik^l_xz}\delta_{m,l}\\
	&(\mathbb{R}_{\pm,z}^{\pm 1})_{m,l}=\pm\gamma_l^{\pm 1} e^{ik^l_xz}\delta_{m,l}\\
	&(\mathbb{H}^j_{\pm,z})_{m,l}=\pm e^{iq^l_xz}J_{m-l}(\alpha_j)\\
	&(\mathbb{L}_{\pm,z}^{j,\pm 1})_{m,l}=\pm \Gamma_l^{\pm 1} e^{iq^l_xz}J_{m-l}(\alpha_j)\\
	&(\mathbb{K}^j_{\pm,z})_{m,l}=\pm e^{iq^l_xz}J_{m-l}(\alpha_j)e^{-i(m-l)\beta}\\
	&(\mathbb{F}_{\pm,z}^{j,\pm 1})_{m,l}=\pm \Gamma_l^{\pm 1} e^{iq^l_z}J_{m-l}(\alpha_j)e^{-i(m-l)\beta}.
\end{align}

We assume the electron propagation direction is from left to right with energy $\varepsilon$ {along $x$-direction}, therefore, the transmission coefficient of the m-$th$ band is written as follows:
\begin{equation}
	t_m=\mathbb{S'}\delta_{m,0}
\end{equation}
with $\mathbb{S'}=\mathbb{S}^{-1}_{1,1}$.
Recall that  $m$ varies from -$N$ to $N$, that is to say
\begin{equation}
	\begin{pmatrix}
		t_{-N}\\
		.\\.\\t_{-1}\\t_0\\t_1\\.\\.\\t_{N}
	\end{pmatrix}=\mathbb{S'}\begin{pmatrix}
		0\\.\\.\\0\\1\\0\\.\\.\\0
	\end{pmatrix}.
\end{equation}
To simplify, we  limit our study to the three first bands: the central band  $\varepsilon$ and the two first side bands  $\varepsilon\pm\varpi$. In this case, we obtain
\begin{align}
	&t_{-1}=\mathbb{S'}[1,2]\\
	&t_{0}=\mathbb{S'}[2,2]\\
	&t_{1}=\mathbb{S'}[3,2].
\end{align}
 The transmission probability can be derived from the current density $J=\Psi^\dagger\sigma_x\Psi$, obtained from the continuity equation. As a result, the incident, reflected, and transmitted densities are given by
\begin{align}
	&J^0_\text{inc}=\gamma_0+\gamma^*_0\\
	&J^m_\text{ref}=r_mr_m^*(\Gamma_m+\Gamma^*_m)\\
	&J^m_\text{tra}=t_mt_m^*(\gamma_m+\gamma^*_m)
\end{align}
where $\gamma_m=s_m\frac{k^m_x+i(k_y-m\omega)}{\sqrt{(k^m_x)^2+(k_y-\varpi)^2}}$ and $\Gamma_m=s_m\frac{q^m_x+i(k_y-m\omega)}{\sqrt{(q^m_x)^2+(k_y-m\varpi)^2}}$.
The transmission corresponding to the $m$-th side-band, is then write as 
\begin{equation}\label{A30}
	T_m=\frac{\cos \theta_m}{\cos \theta_0}|t_m|^2
\end{equation}
where $\cos \theta_m= \frac{k_x^m}{\sqrt{(k_x^m)^2+k_y^2}}$ and $k_x^m=\sqrt{(\varepsilon+m\varpi)^2-k_y^2}$.
Finally, the total transmission is given by the sum over all modes
\begin{equation}\label{A31}
	T=\sum_{m=-N}^{N}T_m.
\end{equation}

\end{document}